\documentclass[11pt,english,twocolumn]{smfart}
\usepackage{mathptmx}
\title{Symmetry classes in random matrix theory}
\author{Martin R.~Zirnbauer}
\date{December 31, 2003}
\address{Institut f\"ur Theoretische Physik, Uni\-versit\"at zu K\"oln,
Z\"ulpicher Str.\ 77, 50937 K\"oln, Germany}
\email{zirn@thp.uni-koeln.de}
\begin{document}
\maketitle

\section{Introduction}

A classification of random-matrix ensembles by symmetries was
first established by Dyson, in an influential 1962 paper with the
title ``the threefold way: alge\-braic structure of symmetry
groups and ensembles in quantum mechanics''. Dyson's threefold way
has since become fundamental to various areas of theoretical
phy\-sics, including the statistical theory of complex many-body
systems, mesoscopic physics, disordered electron systems, and the
field of quantum chaos.

Over the last decade, a number of random-matrix ensembles beyond
Dyson's classification have come to the fore in physics and
mathematics.  On the physics side these emerged from work on the
low-energy Dirac spectrum of quantum chromodynamics, and from the
mesoscopic physics of low-energy quasi-particles in disordered
superconductors.  In the mathematical research area of number
theory, the study of statistical correlations in the values of
Riemann zeta and similar functions has prompted some of the same
generalizations.

In this article, Dyson's fundamental result will be reviewed from
a modern perspective, and the recent extension of Dyson's
threefold way will be motivated and described.  In particular, it
will be explained why symmetry classes are associated with large
families of symmetric spaces.

\section{The framework}\label{sec:frame}

Random matrices have their physical origin in the quantum world,
more precisely in the statistical theory of strongly interacting
many-body systems such as atomic nuclei.  Although random-matrix
theory is nowadays understood to be of relevance to numerous areas
of physics -- see e.g.~Guhr's article in this volume -- quantum
mechanics is still where many of its applications lie. Quantum
mechanics also provides a natural framework in which to classify
random-matrix ensembles.

Following Dyson, the mathematical setting for classification
consists of two pieces of data:
\begin{itemize}
{\item[$\bullet$] A finite-dimensional complex vector space
${\mathcal V}$ with a Hermitian scalar product $\langle \cdot ,
\cdot \rangle$, called a {\it unitary structure} for short. (In
physics applications, ${\mathcal V}$ will usually be the truncated
Hilbert space of a family of quantum Hamiltonian systems.)}
{\item[$\bullet$] On ${\mathcal V}$ there acts a group $G$ of
unitary and anti-unitary operators (the joint symmetry group of
the multi-parameter family of quantum systems).}
\end{itemize}
Given this setup, one is interested in the linear space of
self-adjoint operators on ${\mathcal V}$ -- the Hamiltonians $H$
-- with the property that they commute with $G$. Such a space is
reducible in general, i.e.~the matrix of $H$ decomposes into
blocks. The goal of classification is to enumerate the irreducible
blocks that occur.

\subsection{Symmetry groups}

Basic to classification is the notion of a symmetry group in
quantum Hamiltonian systems, a notion that will now be explained.

In classical mechanics the symmetry group $G_0$ of a Hamiltonian
system is understood to be the group of canonical transformations
that commute with the phase flow of the system. An important
example is the rotation group for systems in a central field.

In passing from classical to quantum mechanics, one replaces the
classical phase space by a quantum mechanical Hilbert space
${\mathcal V}$ and assigns to the symmetry group $G_0$ a
(projective) representation by unitary ${\mathbb C}$-linear
operators on ${\mathcal V}$. Beside the one-parameter continuous
subgroups, whose significance is highlighted by Noether's theorem,
the components of $G_0$ not connected with the identity play an
important role. A prominent example is provided by the operator
for space reflection. Its eigenspaces are the subspaces of states
with positive and negative parity, which reduce the matrix of any
reflection-invariant Hamiltonian to two blocks.

Not all symmetries of a quantum mechanical system are of the
canonical, unitary kind: the prime counterexample is the operation
of inverting the time direction, called time reversal for short.
In classical mechanics this operation reverses the sign of the
symplectic structure of phase space; in quantum mechanics its
algebraic properties reflect the fact that inverting the time
direction, $t \mapsto -t$, amounts to sending ${\rm i} = \sqrt{
-1}$ to $-{\rm i}$.  Indeed, time $t$ enters in the Dirac, Pauli,
or Schr\"odinger equation as ${\rm i} \hbar d/dt$. Therefore, time
reversal is represented in the quantum theory by an
\emph{anti}-unitary operator $T$, which is to say that $T$ is
complex anti-linear:
\begin{displaymath}
    T (z\psi) = {\bar z} \, T \psi \quad (z \in {\mathbb C}, \;
    \psi \in {\mathcal V}) \;,
\end{displaymath}
and preserves the Hermitian scalar product or unitary structure up
to complex conjugation:
\begin{displaymath}
  \left\langle T \psi_1 , T \psi_2 \right\rangle
  = \overline{\left\langle \psi_1 , \psi_2 \right\rangle }
  = \left\langle \psi_2 , \psi_1 \right\rangle \;.
\end{displaymath}
Another operation of this kind is charge conjugation in
relativistic theories such as the Dirac equation.

By the symmetry group $G$ of a quantum mechanical system with
Hamiltonian $H$, one then means the group of all unitary and
anti-unitary transformations $g$ of ${\mathcal V}$ that leave the
Hamiltonian invariant: $g H g^{-1} = H$.  We denote the unitary
subgroup of $G$ by $G_0$, and the set of anti-unitary operators in
$G$ by $G_1$ (not a group). If ${\mathcal V}$ carries extra
structure, as will be the case for some extensions of Dyson's
basic scheme, the action of $G$ on ${\mathcal V}$ has to be
compatible with that structure.
%

The set $G_1$ may be empty. When it is not, the composition of any
two elements of $G_1$ is unitary, so every $g \in G_1$ can be
obtained from a fixed element of $G_1$, say $T$, by right
multiplication with some $U \in G_0$: $g = T U$.  In other words,
when $G_1$ is non-empty the coset space $G / G_0$ consists of
exactly two elements, $G_0$ and $T \cdot G_0 = G_1$.  We shall
assume that $T$ represents some {\it inversion} symmetry such as
time reversal or charge conjugation. $T$ must then be a
(projective) involution, i.e.~$T^2 = z \times {\rm Id}$ with $z$ a
complex number of unit modulus, so that conjugation by $T^2$ is
the identity operation.  Since $T$ is complex anti-linear, the
associative law $T^2 \cdot T = T \cdot T^2$ forces $z$ to be real,
and hence $T^2 = \pm {\rm Id}$.

Finding the total symmetry group of a
%
%
Hamiltonian system
need not always be straightforward,
%
%
but this complication will not be an issue here: we take the
symmetry group $G$ and its action on the Hilbert space ${\mathcal
V}$ as {\it fundamental and given}, and then ask what is the
corresponding {\it symmetry class}, meaning the linear space of
Hamiltonians on ${\mathcal V}$ that commute with $G$.

For technical reasons, we assume the group $G_0$ to be compact;
this is an assumption that covers most (if not all) of the cases
of interest in physics. The non-compact group of space
translations can be incorporated, if necessary, by wrapping the
system around a torus, whereby translations are turned into
compact torus rotations.
%

While the primary objects to classify are the spaces of
Hamiltonians $H$, we shall focus for convenience on the spaces of
{\it time evolutions} $U_t = {\rm e}^{-{\rm i}t H /\hbar}$
instead. This change of focus results in no loss, as the
Hamiltonians can always be retrieved by linearizing in $t$ at $t =
0$.
%

\subsection{Symmetric spaces}

We appropriate a few basic facts from the theory of symmetric
spaces.

Let $M$ be a connected $m$-dimensional Riemannian manifold and $p$
a point of $M$.  In some open subset $N_p$ of a neighborhood of
$p$ there exists a map $s_p : N_p \to N_p$, the {\it geodesic
inversion} with respect to $p$, which sends a point $x \in N_p$
with normal coordinates $(x_1, \ldots, x_m)$ to the point with
normal coordinates $(- x_1, \ldots, - x_m)$. The Riemannian
manifold $M$ is called locally symmetric if the geodesic inversion
is an isometry, and is called {\it globally symmetric} if $s_p$
extends to an isometry $s_p : M \to M$, for all $p \in M$. A
globally symmetric Riemannian manifold is called a symmetric space
for short.

The Riemann curvature tensor of a symmetric space is covariantly
constant, which leads one to distinguish between three cases: the
scalar curvature can be positive, zero, or negative, and the
symmetric space is said to be of compact type, Euclidean type, or
non-compact type, respectively. (In mesoscopic physics each type
plays a role: the first provides us with the scattering matrices
and time evolutions, the second with the Hamiltonians, and the
third with the transfer matrices.) The focus in the current
article will be on compact type, as it is this type that houses
the unitary time evolution operators of quantum mechanics. The
compact symmetric spaces are subdivided into two major subtypes,
both of which occur naturally in the present context, as follows.

\subsection{Type II}

Consider first the case where the anti-unitary component $G_1$ of
the symmetry group is empty, so the data are $({\mathcal V},G)$
with $G = G_0$. Let ${\mathcal U}({\mathcal V})$ denote the group
of all complex linear transformations that leave the structure of
the vector space ${\mathcal V}$ invariant. Thus ${\mathcal U}
({\mathcal V})$ is a group of unitary transformations if
${\mathcal V}$ carries no more than the usual Hermitian scalar
product; and is some subgroup of the unitary group if ${\mathcal
V}$ does have extra structure (as is the case for the Nambu space
of quasi-particle excitations in a superconductor).  The symmetry
group $G_0$, by acting on ${\mathcal V}$ and preserving its
structure, is contained as a subgroup in ${\mathcal U}({\mathcal
V})$.

Let now $H$ be any Hamiltonian with the prescribed symmetries.
Then the time evolution $t \mapsto U_t = {\rm e}^{-{\rm i}t H /
\hbar}$ generated by $H$ is a one-parameter subgroup of ${\mathcal
U}({\mathcal V})$ which commutes with the $G_0$-action. The total
set of transformations $U_t$ that arise in this way is called the
(connected part of the) {\it centralizer} of $G_0$ in ${\mathcal
U}({ \mathcal V})$, and is denoted by $Z$.  This is the ``good''
set of unitary time evolutions -- the set compatible with the
given symmetries of an ensemble of quantum systems.

The centralizer $Z$ is obviously a group: if $U$ and $V$ belong to
$Z$, then so do their inverses and their product.  What can one
say about the structure of the group $Z$?  In essence, this
question was answered by H.~Weyl in his famous treatise ``The
Classical Groups''.

Since $G_0$ is compact by assumption, its group action on
${\mathcal V}$ is completely reducible, and ${\mathcal V}$ is
guaranteed to have an orthogonal vector space decomposition
\begin{displaymath}
    {\mathcal V} = \sum_\lambda {\mathcal V}_\lambda \simeq
    \sum_\lambda V_{\lambda} \otimes \mathbb{C}^{m_\lambda} \;,
\end{displaymath}
where the sum is over (classes of equivalent) irreducible
$G_0$-representations $\lambda$, the $V_\lambda$ are irreducible
representation spaces for the $G_0$-action, and $m_\lambda$ is the
multiplicity of occurrence of the representation type $\lambda$.
($G_0$ acts trivially on $\mathbb{C}^{m_\lambda}$.) The subspaces
${\mathcal V}_\lambda \simeq V_\lambda \otimes \mathbb{C}^{
m_\lambda}$ will be called the $G_0$-{\it isotypic} components of
${\mathcal V}$. For example, if $G_0$ is the rotation group ${\rm
SO}(3)$, the $G_0$-isotypic components of ${\mathcal V}$ are the
subspaces of states with definite total angular momentum, say
$\lambda = 0, 1, 2, \ldots$; $V_\lambda$ then is an ${\rm SO}
(3)$-irreducible representation space of dimension $2\lambda + 1$;
and $m_\lambda$ is the number of times a multiplet of states with
angular momentum $\lambda$ occurs in ${\mathcal V}$.

Now consider any $U \in Z$. Since $U$ commutes with the
$G_0$-action, it does not connect different $G_0$-isotypic
components. (Indeed, in the example of ${\rm SO}(3)$-invariant
dynamics, angular momentum is conserved and transitions between
different angular momentum sectors are impossible.) Thus every
$G_0$-isotypic component ${\mathcal V}_\lambda$ is an invariant
subspace for the action of $Z$ on ${\mathcal V}$, and $Z$
decomposes as $Z = \prod_\lambda Z_\lambda$ with blocks $Z_\lambda
= Z \big|_{{\mathcal V}_\lambda}$. But one can say even more:
because $U \in Z$ commutes with $G_0$ and $V_\lambda$ is
$G_0$-irreducible, $U$ must act like the identity on $V_\lambda$
by Schur's lemma. Therefore, $Z_\lambda$ acts nontrivially only on
the factor $\mathbb{C}^{m_\lambda}$ in ${\mathcal V}_\lambda
\simeq V_\lambda \otimes \mathbb{C}^{m_\lambda}$, so
\begin{displaymath}
    Z = \prod_\lambda Z_\lambda \simeq \prod_\lambda
    {\rm U}(m_\lambda) \quad (\mbox{direct product}) \;,
\end{displaymath}
if ${\mathcal V}$ is a unitary vector space with no extra
structure.  In the presence of extra structure (which, by
compatibility with the $G_0$-action, restricts to every subspace
${\mathcal V}_\lambda$) the factor $Z_\lambda$ is some subgroup of
${\rm U}(m_\lambda)$. In all cases, $Z$ is a direct product of
connected compact Lie groups $Z_\lambda$.

The focus now shifts from $Z$ to any one of the $Z_\lambda$. So we
fix some $M := Z_\lambda$.  Since $M$ is a group, the operation of
taking the inverse, $U \mapsto U^{-1}$, makes sense for all $U \in
M$. Moreover, being a compact Lie group, the manifold $M$ admits a
left- and right-invariant Riemannian structure in which the
inversion $U \mapsto U^{-1}$ is an isometry. By translation one
gets an isometry
\begin{displaymath}
 s_{U_1} : U \mapsto U_1 U^{-1} U_1
\end{displaymath}
for every $U_1 \in M$.  All these maps $s_{U_1}$ are globally
defined, and the restriction of $s_{U_1}$ to some neighborhood of
$U_1$ coincides with the geodesic inversion w.r.t.~$U_1$. Thus $M$
is a symmetric space by the definition given above. Symmetric
spaces of this kind (with group structure) are called type II.

\subsection{Type I}

Consider next the case $G_1 \not= \emptyset$, where some
anti-unitary symmetry $T$ is present.  As before, let $Z$ be the
connected component of the centralizer of $G_0$ in ${\mathcal U}
({\mathcal V})$. Conjugation by $T$,
\begin{displaymath}
    U \mapsto \tau(U) := T U T^{-1} \;,
\end{displaymath}
is an automorphism of ${\mathcal U} ({\mathcal V})$ and, owing to
$T^2 = \pm {\rm Id}$, $\tau$ is involutive. Because $G_0 \subset
G$ is a normal subgroup, $\tau$ restricts to an involutive
automorphism (still denoted by $\tau$) of $Z$. Now recall that $T$
is complex anti-linear and the good Hamiltonians are subject to $T
H T^{-1} = H$. The good time evolutions $U_t = {\rm e}^{-{\rm i}t
H / \hbar}$ clearly satisfy $\tau(U_t) = U_{-t} = U_t^{-1}$. Thus
the good set to consider is
\begin{displaymath}
    {\mathcal M} := \{ U \in Z \, | \, U = \tau(U)^{-1} \} \;.
\end{displaymath}
${\mathcal M}$ is a manifold, but in general is not a Lie group.

Further details depend on what $\tau$ does with the factorization
$Z = \prod_\lambda Z_\lambda$. If ${\mathcal V}_\lambda$ is a
$G_0$-isotypic component of ${\mathcal V}$, then so is $T
{\mathcal V}_\lambda$, since $T$ normalizes $G_0$. Thus $T
{\mathcal V}_\lambda = {\mathcal V}_{\tilde\lambda}$ for some
representation type $\tilde\lambda$.  If $\tilde\lambda \not=
\lambda$, the involutive automorphism $\tau$ just relates $U \in
Z_\lambda$ with $\tau(U) \in Z_{\tilde\lambda}$, whence no intrinsic
constraint on $Z_\lambda$ results, and the time evolutions $\big(
U, \tau(U)^{-1} \big) \in Z_\lambda \times Z_{\tilde\lambda}$
constitute a type-II symmetric space, as before.
%

A novel situation occurs when $\tilde\lambda = \lambda$, in which
case $\tau$ maps the group $Z_\lambda$ onto itself.  Let therefore
$\tilde\lambda = \lambda$, put $K \equiv Z_\lambda$ for short, and
consider
\begin{displaymath}
    M := \{ U \in K \, | \, U = \tau(U)^{-1} \} \;.
\end{displaymath}
Note that if two elements $p, p_0$ of $K$ are in $M$, then so is
the product $p_0 p^{-1} p_0$.  The group $K$ acts on $M \subset K$
by
\begin{displaymath}
    k \cdot U = k\,  U \tau(k)^{-1} \quad (k \in K) \;,
\end{displaymath}
and this group action is transitive, i.e.~every $U \in M$ can be
written as $U = k \tau(k)^{-1}$ with some $k \in K$. (Finding $k$
for a given $U$ is like taking a square root, which is possible
since $\exp : {\rm Lie}\, K \to K$ is surjective.) There exists a
$K$-invariant Riemannian structure for $M$ such that for all $p_0
\in M$ the mapping $s_{p_0} : M \to M$ defined by
\begin{displaymath}
    s_{p_0}(p) = p_0 p^{-1} p_0 \;,
\end{displaymath}
is the geodesic inversion w.r.t.~$p_0 \in M$.  Thus in this
natural geometry $M$ is a globally symmetric Riemannian manifold
and hence a symmetric space.  The present kind of symmetric space
is called type I. If $K_\tau$ is the set of fixed points of $\tau$
in $K$, the symmetric space $M$ is analytically diffeomorphic to
the coset space $K / K_\tau$ by
\begin{displaymath}
K / K_\tau \to M \subset K \;, \quad U K_\tau \mapsto U
\tau(U)^{-1} \;,
\end{displaymath}
which is called the {\it Cartan embedding} of $K / K_\tau$ into
$K$.

In summary, the solution to the problem of finding the unitary
time evolution operators that are compatible with a given symmetry
group and structure of Hilbert space, is always a symmetric space.
This is a valuable insight, as symmetric spaces are rigid objects
%
%
and have been completely classified by Cartan.

If we keep the dimension of $\mathcal{V}$ {\it variable}, the
symmetric spaces that occur must be those of a {\it large family}.

\section{Dyson's threefold way}

Recall the goal: given a Hilbert space ${\mathcal V}$ and a
symmetry group $G$ acting on it, one wants to classify the
(irreducible) spaces of time evolution operators $U$ that are
``compatible'' with $G$, meaning
\begin{displaymath}
U = g_0^{\vphantom{-1}} U g_0^{-1} = g_1^{\vphantom{-1}} U^{-1}
g_1^{-1} \quad (\mbox{for all } g_\sigma \in G_\sigma ) \;.
\end{displaymath}
As we have seen, the spaces that arise in this way are symmetric
spaces of type I or II depending on the nature of the time
reversal (or other anti-unitary symmetry) $T$.

An even stronger statement can be made when more information about
the Hilbert space ${\mathcal V}$ is specified.  In Dyson's
classification, the Hermitian scalar product of ${\mathcal V}$ is
assumed to be the \emph{only} invariant structure that exists on
${\mathcal V}$.  With that assumption, only three large families
of symmetric spaces arise; these correspond to what we call the
{\it Wigner-Dyson symmetry classes}.
%

\subsection{Class $A$}

Recall that in Dyson's case, the connected part of the centralizer
of $G_0$ in ${\mathcal U}({\mathcal V})$ is a direct product of
unitary groups, each factor being associated with one
$G_0$-isotypic component ${\mathcal V}_\lambda$ of ${\mathcal V}$.
The type-II situation occurs when the set $G_1$ of anti-unitary
symmetries is either empty or else exchanges different ${\mathcal
V}_\lambda$. In both cases, the set of good time evolution
operators restricted to one $G_0$-isotypic component ${\mathcal
V}_\lambda$ is a unitary group ${\rm U}(m_\lambda)$, with
$m_\lambda$ being the multiplicity of the irreducible
$G_0$-representation $\lambda$ in ${\mathcal V}_ \lambda$.

The unitary groups ${\rm U}(N = m_\lambda)$ or to be precise,
their simple parts ${\rm SU}(N)$, are called type-II symmetric
spaces of the $A$ family or $A$ series -- hence the name class
$A$.  The Hamiltonians $H$, the generators of time evolutions $U_t
= {\rm e}^{-{\rm i}t H / \hbar}$, in this class are represented by
complex Hermitian $N \times N$ matrices. By putting a ${\rm U}
(N)$-invariant Gaussian probability measure
\begin{displaymath}
    \exp\left( - {\rm Tr}\, H^2 /2\sigma^2 \right) dH
    \quad (\sigma \in \mathbb{R})
\end{displaymath}
on that space, one gets what is called the GUE -- the Gaussian
Unitary Ensemble -- which defines the Wigner-Dyson {\it
universality class} of unitary symmetry.
%

\subsection{Classes $A$I and $A$II}

Consider next the case $G_1 \not= \emptyset$, with anti-unitary
generator $T$.  Let ${\mathcal V}_\lambda$ be any $G_0$-isotypic
component which is invariant under $T$ (the type-I situation). The
mapping $U \mapsto T U T^{-1} = \tau(U)$ then is an automorphism
of the groups ${\rm U}({\mathcal V}_\lambda)$, $G_0$ and $K =
Z_\lambda \simeq {\rm U}(m_\lambda)$. If $K_\tau$ is the subgroup
of fixed points of $\tau$ in $K$, the space of good time
evolutions can be identified with the symmetric space $K / K_\tau$
by the Cartan embedding. The task is to determine $K_\tau$. As was
emphasized by Dyson, the answer for $K_\tau$ does not follow from
any single piece of data, but is determined by the combination of
{\bf three} anti-unitary involutions on ${\mathcal V}_\lambda$.

The first of these is the standard operation of taking the complex
conjugate (w.r.t.~the complex structure of ${\mathcal
V}_\lambda$), denoted by $\psi \mapsto \bar \psi$ as usual. Recall
that the unitary vector space ${\mathcal V}_\lambda$ decomposes as
an orthogonal sum of $m_\lambda$ identical copies of an
irreducible representation space $V_\lambda$ for the compact group
$G_0$:
\begin{equation}\label{zerlegung}
    {\mathcal V}_\lambda = V_\lambda^{(1)} \oplus V_\lambda^{(2)}
    \oplus \ldots \oplus V_\lambda^{(m_\lambda)} \simeq
    V_\lambda \otimes \mathbb{C}^{m_\lambda} \;.
\end{equation}
From the multitude of such decompositions, we select one that is
invariant under complex conjugation: if $\psi$ lies in a subspace
$V_\lambda^{(j)}$, so does $\bar\psi$.

Next we look at the operation $g_0 \mapsto \bar g_0$ on any one of
the $G_0$-irreducible subspaces of ${\mathcal V}_\lambda$, say
$V_\lambda \equiv V_\lambda^{(1)}$.  Since $\lambda = \bar
\lambda$ by assumption, the $G_0$-action on $V_\lambda$ is
unitarily equivalent to its complex conjugate.  Thus there exists
some unitary transformation $s \in {\rm U}(V_\lambda)$ such that
\begin{displaymath}
    \bar g_0 = s^{-1} g_0 s
\end{displaymath}
holds for every $g_0 \in G_0$.  Given $s$, one defines an
anti-unitary operator $S$ on ${\mathcal V}_\lambda$ by
\begin{displaymath}
    S \big( v^{(1)} + \ldots + v^{(m_\lambda)} \big) = s
    \overline{v^{(1)}} + \ldots + s \overline{v^{(m_\lambda)}} \;,
\end{displaymath}
where the decomposition (\ref{zerlegung}) is invoked.  By
construction, $S$ commutes with the action of $G_0$.  Since the
$G_0$-action on $V_\lambda$ is irreducible, Schur's lemma applied
to $S^2$ forces $S$ to be a projective involution, and
associativity ($S^2 \cdot S = S \cdot S^2$) results in $S^2 = \pm
1_{\mathcal{V}_\lambda}$.

Recall next that $g_0 \mapsto T g_0 T^{-1}$ is an automorphism of
$G_0$, and remains so when the $G_0$-action is restricted to
${\mathcal V}_\lambda$. Since all of the $G_0$-representations in
the isotypic component ${\mathcal V}_\lambda$ are equivalent,
there exists a unitary transformation $R \in {\rm U}({\mathcal
V}_\lambda)$ such that
\begin{displaymath}
    T g_0 T^{-1} = R^{-1} g_0 R \quad (\mbox{for all } g_0 \in G_0)
\end{displaymath}
holds as an operator identity on ${\mathcal V}_\lambda$. Note that
the composition $R T$ intertwines $G_0$-actions: $R T g_0 = g_0 R
T$, but changes the complex structure of ${\mathcal V}_\lambda$
(by anti-linearity of $T$).  A better object to consider is $R T
\circ S$ which, being composed of two anti-unitary operators, is
unitary. $RTS$ commutes with the $G_0$-action and thus lies in the
centralizer $K$. Using it, one defines another anti-unitary
operator $T^\prime$ on ${\mathcal V}_\lambda$ and an automorphism
$\tau^\prime$ of ${\rm U}({\mathcal V}_\lambda)$ by
\begin{displaymath}
    T^\prime \psi = RTS \bar\psi \;, \quad \tau^\prime(k) =
    T^\prime k {T^\prime}^{-1} \;.
\end{displaymath}
$T^\prime$ determines a complex bilinear form $Q$ on ${\mathcal
V}_\lambda$ by
\begin{displaymath}
    Q(\psi_1, \psi_2) = \left\langle T^\prime \psi_1 , \psi_2
    \right\rangle_{{\mathcal V}_\lambda} \quad (\mbox{for all }
    \psi_1, \psi_2 \in {\mathcal V}_\lambda) \;.
\end{displaymath}

The remaining steps toward identifying $K_\tau$ depend on the
nature of $R$. Consider first the easy case where the automorphism
$\tau$ of $G_0$ is inner, i.e.~$R \in G_0$.  Then $k = \tau(k)$
for $k \in K$ is equivalent to $k = RT k (RT)^{-1}$, which in turn
amounts to $k = RTS \, \bar{k} (RTS)^{-1} = \tau^\prime (k)$.
Hence another description of $K_\tau$ is to say that its elements
$k$ are the unitary transformations of ${\mathcal V}_\lambda$ that
centralize $G_0$ and leave the pairing $Q$ invariant (the latter
follows from $T^\prime k = k T^\prime$ and invariance of $\langle
\cdot , \cdot \rangle_{\mathcal{V}_\lambda}$).

By iterating $k = \tau^\prime(k) = {\tau^\prime}^2(k)$ one infers
that ${T^\prime}^2$ commutes with the $K_\tau$-action on
${\mathcal V}_\lambda$. But ${T^\prime}^2$ also commutes with
$T^\prime$ and with the $G_0$-action, which implies ${T^\prime}^2
= \epsilon \times 1_{\mathcal{V}_\lambda}$ with $\epsilon = \pm
1$, by standard reasoning. From
\begin{displaymath}
    Q(\psi_1 , \psi_2) = \overline{\big\langle {T^\prime}^2
    \psi_1 , T^\prime \psi_2 \big\rangle}_{\mathcal{V}_\lambda}
    = \epsilon \, Q(\psi_2 , \psi_1)
\end{displaymath}
one sees that the pairing $Q$ is symmetric for $\epsilon = +1$,
and skew for $\epsilon = -1$.  The Lie group $K_\tau$ is now
easily identified. Since the unitary operator $RTS$ is an element
of $K \simeq {\rm U}(m_\lambda)$, $Q$ restricts to a pairing on
the factor $\mathbb{C}^{m_\lambda}$ in the decomposition
${\mathcal V}_\lambda \simeq V_\lambda \otimes \mathbb{C}^{
m_\lambda}$. Thus for $\epsilon = +1$, $K_\tau$ can be viewed as a
subgroup of ${\rm U}(m_\lambda)$ that preserves a symmetric
pairing (or orthogonal structure) on $\mathbb{C}^{ m_\lambda}$;
consequently $K_\tau \simeq {\rm O} (m_\lambda)$. For $\epsilon =
-1$, the multiplicity $m_\lambda$ must be even, and $K_\tau$
preserves a skew pairing (or symplectic structure); in that case
$K_\tau \simeq {\rm USp} (m_\lambda)$, the unitary symplectic
group.

In the general case ($R \notin G_0$) drawing these conclusions is
more difficult, and one must exploit the rigidity of the symmetric
space $K / K_\tau$ under deformations of the automorphism $\tau(U)
= T U T^{-1}$.  Actually, the method used by Dyson to handle the
general case is quite different: Dyson chooses to regard
${\mathcal V}_\lambda$ as a \emph{real} vector space (with complex
structure) and expresses the group action of $G$ by orthogonal
matrices. In this real setup he then exploits a deep theorem of
Weyl on the structure of group algebras and their commutator
algebras, which leads him to the conclusion that the above two
possibilities are in fact the only ones that can occur in the
present type-I situation. Thus there is a dichotomy for the sets
of good time evolutions $M \simeq K / K_\tau$:
\begin{displaymath}
\begin{array}{lll}
    \mbox{Class $A$I}: \hspace{0.2cm} &K/K_\tau \simeq {\rm U}(N)
    / {\rm O}(N) \hspace{0.2cm} &(N = m_\lambda) \;, \\
    \mbox{Class $A$II}: &K/K_\tau \simeq {\rm U}(2N)/{\rm USp}(2N)
    &(2N = m_\lambda) \;.
\end{array}
\end{displaymath}
Again we are referring to symmetric spaces by the names they -- or
rather their simple parts ${\rm SU}(N)/{\rm SO}(N)$ and ${\rm SU}
(2N) / {\rm USp}(2N)$ -- have in the Cartan classification. Be
warned that Weyl's theorem by itself does not allow to decide
between the alternatives $A$I or $A$II for a given data set
$({\mathcal V}_\lambda, G_0, T)$ (rather, to do so you must
determine the ``Wigner type'' of the $G$-representation on
${\mathcal V}_\lambda$).

In the case where the $G_0$-automorphism $\tau$ is inner (which
actually covers most of the known examples of physical interest)
Dyson's reasoning is basically identical to the one reviewed
above. There, as we have seen, the dichotomy is ruled by the
number $\epsilon$ computed from $(T^\prime)^2 = RTS \, \overline{
RTS} = \epsilon \times 1_{\mathcal{V}_\lambda}$.  Important
examples are provided by physical systems with spin-rotation
symmetry, $G_0 = {\rm SU}(2)$, and time-reversal symmetry. The
physical operation of time reversal, $T$, commutes with spin
rotations, so $\tau$ is inner here with $R = 1$. On states with
spin $|S|$, one has $T^2 = (-1)^{2|S|}$ and $S^2 = (-1)^{2|S|}$,
which gives ${T^\prime}^2 = +1$ in all cases. Thus time-reversal
invariant systems with no symmetries other than energy and spin
are always class $A$I. By breaking spin-rotation symmetry ($G_0 =
\{ {\rm Id} \}$, so $S^2 = 1$) while maintaining $T$-symmetry for
states with half-integer spin (say single electrons, which carry
spin $|S| = 1/2$), one gets ${T^\prime}^2 = T^2 = -1$, thereby
realizing class $A$II.

The Hamiltonians $H$, obtained by passing to the tangent space of
$K / K_\tau$ at unity, are represented by Hermitian matrices with
entries that are real numbers (class $A$I) or real quaternions
(class $A$II). If you put $K_\tau$-invari\-ant Gaussian
probability measures on these spaces, you get the Wigner-Dyson
universality classes of orthogonal resp.~symplectic symmetry.  In
mesoscopic physics these are realized in disordered metals with
time-reversal invariance (absence of magnetic fields and magnetic
impurities). Spin-rotation symmetry is broken by strong spin-orbit
scatterers such as gold impurities.

\section{Disordered superconductors}
\label{sec:superconds}

When Dirac first wrote down his famous equation in 1928, he
assumed that he was writing an equation for the
\emph{wavefunction} of the electron. Later, because of the
instability caused by negative-energy solutions, the Dirac
equation was reinterpreted (via second quantization) as an
equation for the fermio\-nic \emph{field operators} of a quantum
field theory. A similar change of viewpoint is carried out in
reverse in the Hartree-Fock-Bogoliubov mean field description of
quasi-particle excitations in superconductors. There, one starts
from the equations of motion for linear superpositions of the
electron creation and annihilation operators, and reinterprets
them as a unitary quantum dynamics for what might be called the
quasi-particle ``wavefunction''.

In both cases -- the Dirac equation and the quasi-particle
dynamics of a superconductor -- there enters a structure not
present in the standard quantum mechanics underlying Dyson's
classification: the field operators for fermionic particles are
subject to a set of requirements called the {\it canonical
anti-commutation relations}, and these are preserved by the
quantum dynamics. Therefore, whenever second quantization is
undone (assuming it \emph{can} be undone) to return from field
operators to wavefunctions, the wavefunction dynamics is required
to preserve some extra structure.  This puts a linear constraint
on the allowed Hamiltonians $H$. For our purposes, the best
viewpoint to take is to attribute the extra invariant structure to
the Hilbert space ${\mathcal V}$, thereby turning it into a Nambu
space.

\subsection{Nambu space}


Starting from the standard for\-malism of second quantization,
consider a set of single-particle creation and annihilation
operators $c_\alpha^\dagger$ and $c_\alpha^{\vphantom{\dagger}}$,
where $\alpha = 1, \ldots, N$ labels single-particle states that
are orthogonal to each other. Such operators are subject to the
canonical anti-commutation relations
\begin{eqnarray}\label{CAR}
    &&\hspace{1cm} c_\alpha^\dagger c_\beta^{\vphantom{\dagger}}
    + c_\beta^{ \vphantom{\dagger}} c_\alpha^\dagger =
    \delta_{\alpha\beta} \;, \\ &&c_\alpha^\dagger c_\beta^\dagger
    + c_\beta^\dagger c_\alpha^\dagger = 0 = c_\alpha c_\beta +
    c_\beta c_\alpha \nonumber \;.
\end{eqnarray}
When written in terms of $c_\alpha + c_\alpha^\dagger$ and ${\rm
i} (c_\alpha - c_\alpha^\dagger)$, these become the defining
relations of a Clifford algebra. Field operators are linear
combinations $\psi = \sum_\alpha \big( u_\alpha^{\vphantom{
\dagger}} c_\alpha^\dagger + v_\alpha c_\alpha \big)$ with complex
coefficients $u_\alpha$ and $v_\alpha$.

Now take $H$ to be some Hamiltonian which is quadratic in the
creation and annihilation operators:
\begin{displaymath}
    H = \sum_{\alpha \beta} A_{\alpha\beta} c_\alpha^\dagger
    c_\beta + \frac{1}{2} \sum_{\alpha\beta} \left( B_{\alpha
    \beta} c_\alpha^\dagger c_\beta^\dagger + \bar
    B_{\alpha\beta} c_\beta c_\alpha \right) \;,
\end{displaymath}
and let $H$ act on field operators $\psi$ by the commutator: $H
\cdot \psi \equiv [ H , \psi ]$. The time evolution of $\psi$ is
then determined by the equation
\begin{equation}\label{evolution}
  \frac{d \psi}{dt} = - \frac{\rm i}{\hbar} H \cdot \psi \;,
\end{equation}
which integrates to $\psi(t) = {\rm e}^{-{\rm i} t H / \hbar}
\cdot \psi(0)$, and is easily verified to preserve the relations
(\ref{CAR}).

The dynamical equation (\ref{evolution}) is equivalent to a system
of linear differential equations for the amplitudes $u_\alpha$ and
$v_\alpha$.  If these are assembled into vectors, and the
$A_{\alpha\beta}$ and $B_{\alpha\beta}$ into matrices, equation
(\ref{evolution}) becomes
\begin{displaymath}
    \frac{d}{dt} \begin{pmatrix} {\bf u}\\ {\bf v}\end{pmatrix} = -
    \frac{\rm i}{\hbar} \begin{pmatrix} A &B \\ -\bar B &-\bar A
    \end{pmatrix} \begin{pmatrix} {\bf u}\\ {\bf v}\end{pmatrix} \;.
\end{displaymath}
The Hamiltonian matrix on the right-hand side has some special
properties due to $B_{\alpha\beta} = - B_{\beta \alpha}$ (from
$c_\alpha c_\beta = - c_\beta c_\alpha$) and $A_{\alpha\beta} =
{\bar A}_{\beta\alpha}$ (from $H$ being self-adjoint as an
operator in Fock space).
%
%
To keep track of these properties while imposing some unitary and
anti-unitary symmetries, it is best to put everything in invariant
form.

Let $V$ be the complex vector space of annihilation operators $v =
\sum_\alpha v_\alpha c_\alpha$, and view the creation operators $u
= \sum_\alpha u_\alpha^{\vphantom{\dagger}} c_\alpha^\dagger$ as
lying in the dual vector space $V^\ast$. The field operators $\psi
= u + v$ then are elements of the direct sum $V^\ast \oplus V =:
{\mathcal V}$, called {\it Nambu space}.  On ${\mathcal V}$ there
exists a canonical unitary structure
%
%
expressed by
\begin{displaymath}
    \big\langle u_1 + v_1 , u_2 + v_2 \big\rangle =
    \sum_\alpha \left( \bar u_{1 \alpha} u_{2 \alpha} + \bar
    v_{1 \alpha} v_{2 \alpha} \right) \;.
\end{displaymath}
%
%
A second canonical structure on ${\mathcal V}$ is given by the
symmetric $\mathbb{C}$-bilinear form
\begin{displaymath}
    \{ u_1 + v_1 , u_2 + v_2 \} = \sum_\alpha \left( u_{1 \alpha}
    v_{2 \alpha} + u_{2 \alpha} v_{1 \alpha} \right) \;.
\end{displaymath}
Note that $\{ \psi_1 , \psi_2 \}$ agrees with the anti-commutator
of the field operators, $\psi_1 \psi_2 + \psi_2 \psi_1$, by the
relations (\ref{CAR}).

Now recall that the quantum dynamics is determined by a
Hamiltonian $H$ that acts on $\psi$ by the commutator $H \cdot
\psi = [H , \psi]$.  The one-parameter groups $t \mapsto {\rm
e}^{-{\rm i}tH/\hbar}$ generated by this action (the time
evolutions) preserve the symmetric pairing:
\begin{displaymath}
    \{ \psi_1 , \psi_2 \} = \{ {\rm e}^{-{\rm i} t H/\hbar} \cdot
    \psi_1 , {\rm e}^{-{\rm i}t H / \hbar} \cdot \psi_2 \} \;,
\end{displaymath}
since the anti-commutation relations (\ref{CAR}) do not change
with time. They also preserve the unitary structure,
\begin{displaymath}
    \big\langle \psi_1 , \psi_2 \big\rangle = \big\langle
    {\rm e}^{-{\rm i} t H / \hbar} \cdot \psi_1 , {\rm e}^{-
    {\rm i}t H / \hbar} \cdot \psi_2 \big\rangle \;,
\end{displaymath}
because probability in Nambu space is conserved. (Physically
speaking, this holds true as long as $H$ is quadratic,
i.e.~many-body interactions are negligible.)

One can now pose Dyson's question again: given Nambu space
${\mathcal V}$ and a symmetry group $G$ acting on it, what is the
set of time evolution operators that preserve the structure of
${\mathcal V}$ and are compatible with $G$?  From Section
\ref{sec:frame} we know the answer to be some symmetric space, but
which are the symmetric spaces that occur?

\subsection{Class $D$}

Consider a superconductor with no symmetries in its quasi-particle
dynamics, so $G = \{ {\rm Id} \}$. (A concrete example would be a
disordered spin-triplet superconductor in the vortex phase). The
time evolutions $U = {\rm e}^{-{\rm i}t H / \hbar}$ are then
constrained only by invariance of the unitary structure and the
symmetric pairing $\{ \cdot , \cdot \}$ of Nambu space. These two
structures are consistent; they are related by {\it particle-hole
conjugation} $C$:
\begin{displaymath}
    \{ \psi_1 , \psi_2 \} = \big\langle C \psi_1 , \psi_2
    \big\rangle \;,
\end{displaymath}
which is an anti-unitary operator with square $C^2 = + {\rm Id}$.
The condition $\{ \psi_1 , \psi_2 \} = \{ U \psi_1 , U \psi_2 \}$
(invariance of an orthogonal structure) selects a complex
orthogonal group, and imposing unitarity yields a real subgroup
${\rm SO}({\mathcal V}) \simeq {\rm SO}(4N)$ -- a symmetric space
of the $D$ family.

Since the time evolutions are a real orthogonal group, there
exists a basis of ${\mathcal V}$ (called {\it Majorana fermions}
in physics) in which the matrix of ${\rm i}H \in \mathfrak{so}(
{\mathcal V})$ is real skew, and that of $H$ imaginary skew. The
simplest random matrix model for class $D$, the ${\rm SO}
$-invariant Gaussian ensemble of imaginary skew matrices, is
analyzed in the second edition of Mehta's book.  From the
expressions given by Mehta it is seen that the level correlation
functions at high energy coincide with those of the Wigner-Dyson
universality class of unitary symmetry. The level correlations at
low energy, however, show different behavior defining a separate
universality class.  This universal behavior at low energies has
immediate physical relevance, as it is precisely the low-energy
quasi-particles that determine the thermal transport properties of
the superconductor at low temperatures.

\subsection{Class $D$III}

Let now magnetic fields and magnetic impurities be absent, so that
time reversal $T$ is a symmetry of the quasi-particle system: $G =
\{ {\rm Id}, T \}$.  Following Section \ref{sec:frame}, the set of
good time evolutions is $M \simeq K / K_\tau$ with $K = {\rm
SO}({\mathcal V})$ and $K_\tau$ the set of fixed points of $U
\mapsto \tau(U) = T U T^{-1}$ in $K$.  What is $K_\tau$?

The time-reversal operator has square $T^2 = - {\rm Id}$ (for
particles with spin $1/2$), and commutes with particle-hole
conjugation $C$, which makes $Q := {\rm i}CT$ a useful operator to
consider. Since $C$ by definition commutes with the action of $K$,
and hence also with that of $K_\tau$, the subgroup $K_\tau$ has an
equivalent description as
\begin{displaymath}
    K_\tau = \{ k \in {\rm U}({\mathcal V}) \, | \, k = Q k Q^{-1}
    = \tau(k) \} \;.
\end{displaymath}
The operator $Q$ is easily seen to have the following
pro\-perties: (i) $Q$ is unitary, (ii) $Q^2 = {\rm Id}$, and (iii)
${\rm Tr}_{ \mathcal V} Q = 0$.  Consequently $Q$ possesses two
eigenspaces ${\mathcal V}_{\pm}$ of equal dimension, and the
condition $k = Q k Q^{-1}$ fixes a subgroup ${\rm U}({\mathcal V}
_+) \times {\rm U}({\mathcal V}_-)$ of ${\rm U}({\mathcal V})$.
Since $Q$ contains a factor ${\rm i} = \sqrt{-1}$ in its
definition, it anti-commutes with the anti-linear operator $T$.
Therefore the automorphism $\tau$ exchanges ${\rm U}({\mathcal
V}_+)$ with ${\rm U}({\mathcal V}_-)$, and the fixed point set
$K_\tau$ is the same as ${\rm U}({\mathcal V}_+) \simeq {\rm
U}(2N)$.  Thus
\begin{displaymath}
    M \simeq K / K_\tau \simeq {\rm SO}(4N) / {\rm U}(2N) \;,
\end{displaymath}
a symmetric space in the $D$III family.  Note that for particles
with spin $1/2$ the dimension of $\mathcal{V}_+$ has to be even.

By realizing the algebra of involutions $C, T$ as $C \psi = ({\rm
i}\sigma_x \otimes 1_{2N}) \bar\psi$ and $T \psi = ({\rm
i}\sigma_y \otimes 1_{2N}) \bar\psi$, the Hamiltonians $H$ in
class $D$III are brought into the standard form
\begin{displaymath}
    H = \begin{pmatrix} 0 &Z\\ -\bar Z &0 \end{pmatrix} \;,
\end{displaymath}
where the $2N \times 2N$ matrix $Z$ is complex and skew.

\subsection{Class $C$}

Next let the spin of the quasi-particles be conserved, as is the
case for a spin-singlet superconductor with no spin-orbit
scatterers present, and let time-reversal invariance be broken by
a magnetic field. The symmetry group of the quasi-particle system
then is the spin-rotation group: $G = G_0 = {\rm Spin} (3) = {\rm
SU}(2)$.

Nambu space ${\mathcal V}$ can be arranged to be a tensor product
${\mathcal V} = {\mathcal W} \otimes \mathbb{C}^2$ so that $G_0$
acts trivially on ${\mathcal W}$ and by the spinor representation
on the spinor space $\mathbb{C}^2$. Since two spinors combine to
give a scalar, the latter comes with a skew-symmetric form
$\varepsilon : {\mathbb C}^2 \times {\mathbb C}^2 \to \mathbb{C}$.
In a suitable basis, the anti-commutation relations (\ref{CAR})
factor on particle-hole and spin indices. The symmetric bilinear
form $\{ \cdot , \cdot \}$ of ${\mathcal V}$ correspondingly
factors under the tensor product decomposition ${\mathcal V} =
{\mathcal W} \otimes \mathbb{C}^2$ as
\begin{displaymath}
    \{ w_1 \otimes s_1 , w_2 \otimes s_2 \} = [ w_1 , w_2 ] \times
    \varepsilon( s_1 , s_2 ) \;,
\end{displaymath}
where $[ \cdot , \cdot ]$ is a skew bilinear form on ${\mathcal
W}$, giving ${\mathcal W}$ the structure of a symplectic vector
space.

The good set $M$ now consists of the time evolutions that, in
addition to preserving the structure of Nambu space, commute with
the spin-rotation group ${\rm SU}(2)$:
\begin{displaymath}
    M = \{ U \in {\rm U}({\mathcal V}) | U C = C U ,
    \forall R \in {\rm SU}(2): R U = U R \} \;.
\end{displaymath}
By the last condition, all time evolutions act trivially on the
factor $\mathbb{C}^2$. The condition $U C = C U$, which expresses
invariance of the orthogonal structure of ${\mathcal V}$, then
implies that time evolutions preserve the symplectic pairing of
${\mathcal W}$. Time evolutions therefore are unitary symplectic
transformations of ${\mathcal W}$, hence $M = {\rm USp}({\mathcal
W}) \simeq {\rm USp}(2N)$ -- a symmetric space of the $C$ family.
The Hamiltonian matrices in class $C$ have the standard form
\begin{displaymath}
    H = \begin{pmatrix} A &B \\ \bar B &-\bar A \end{pmatrix}
\end{displaymath}
with $A$ being Hermitian and $B$ complex and symmetric.

\subsection{Class $C$I}\label{sec:classCI}

The next class is obtained by taking the time reversal $T$ as well
as the spin rotations $R \in {\rm SU}(2)$ to be symmetries of the
quasi-particle system.

By arguments that should be familiar by now, the set of good time
evolutions is a symmetric space $M \simeq K / K_\tau$ with $K =
{\rm USp}({\mathcal W})$ and $K_\tau$ the set of fixed points of
$\tau$ in $K$.  Once again, the question to be answered is: what's
$K_\tau$?  The situation here is very similar to the one for class
$D$III, with ${\mathcal W}$ and ${\rm USp}({\mathcal W})$ taking
the roles of ${\mathcal V}$ and ${\rm SO}({\mathcal V})$.  By
adapting the previous argument to the present case, one shows that
$K_\tau$ is the same as ${\rm U}({\mathcal W}_+) \simeq {\rm U}
(N)$, where ${\mathcal W}_+$ is the positive eigenspace of $Q =
{\rm i}CT$ viewed as a unitary operator on ${\mathcal W}$. Thus
\begin{displaymath}
    M \simeq K / K_\tau \simeq {\rm USp}(2N) / {\rm U}(N) \;.
\end{displaymath}
The standard form of the Hamiltonian matrices here is
\begin{displaymath}
    H = \begin{pmatrix} 0 &Z\\ \bar Z &0 \end{pmatrix}
\end{displaymath}
with the $N \times N$ matrix $Z$ being complex and symmetric.

\section{Dirac fermions: the chiral classes}

Three large families of symmetric spaces remain to be implemented.
Although these, too, occur in mesoscopic physics, their most
natural realization is by 4d Dirac ferm\-ions in a random gauge
field background.

Consider the Lagrangian $L$ for the Euclidean space-time version
of quantum chromodynamics with $N_c \ge 3$ colors of quarks
coupled to an ${\rm SU}(N_c)$ gauge field $A_\mu$:
\begin{displaymath}
    L = {\rm i} \bar\psi\, \gamma^\mu (\partial_\mu - A_\mu) \psi
    + {\rm i}m \bar\psi \psi\;.
\end{displaymath}
The massless Dirac operator $D = {\rm i} \gamma^\mu (\partial_\mu
- A_\mu)$ anti-com\-mutes with $\gamma_5 = \gamma^0 \gamma^1
\gamma^2 \gamma^3$.  Therefore, in a basis of eigenstates of
$\gamma_5$ the matrix of $D$ takes the form
\begin{equation}\label{classAIII}
    D = \begin{pmatrix} 0 &Z \\ Z^\dagger &0 \end{pmatrix} \;.
\end{equation}
If the gauge field carries topological charge $\nu \in
\mathbb{Z}$, the Dirac operator $D$ has at least $|\nu|$ zero
modes by the index theorem.  To make a simple model of the
challenging situation where $A_\mu$ is distributed according to
Yang-Mills measure, one takes the matrices $Z$ to be complex
rectangular, of size $p \times q$ with $p - q = \nu$, and puts a
Gauss measure on that space. This random-matrix model for $D$
captures the universal features of the QCD Dirac spectrum in the
massless limit.

The exponential of the truncated Dirac operator, ${\rm e}^{{\rm
i}t D}$ (where $t$ is not the time), lies in a space equivalent to
${\rm U}(p+q) / {\rm U}(p) \times {\rm U}(q)$ -- a symmetric space
of the $A$III family.  We therefore say that the universal
behavior of the QCD Dirac spectrum is that of symmetry class
$A$III.

But hold on! Why are we entitled to speak of a {\it symmetry
class} here? By definition, symmetries always {\it commute} with
the Hamiltonian, never do they anti-com\-mute!  (The relation $D =
- \gamma_5 D\, \gamma_5$ is {\it not a symmetry} in the sense of
Dyson, nor is it a symmetry in our sense.)

\subsection{Class $A$III}

To incorporate the massless QCD Dirac operator into the present
classification scheme, we adapt it to the Nambu space setting.
This is done by reorganizing the $4$-component Dirac spinor
$\psi$, $\bar\psi$ as an $8$-component Majorana spinor $\Psi$, to
write
\begin{displaymath}
    L_{m = 0} = \frac{\rm i}{2} \Psi \, \Gamma^\mu (\partial_\mu -
    \mathcal{A}_\mu ) \Psi \;.
\end{displaymath}
The $8 \times 8$ matrices $\Gamma^\mu$ are real symmetric besides
satisfying the Clifford relations $\Gamma^\mu \Gamma^\nu +
\Gamma^\nu \Gamma^\mu = 2 \delta^{\mu\nu}$. A possible
tensor-product realization is
\begin{eqnarray*}
    &&\Gamma^0 = \,\,\,\, 1 \otimes \sigma_z \otimes 1 \;, \quad
    \Gamma^1 = \sigma_x \otimes \sigma_y \otimes \sigma_y \;, \\
    &&\Gamma^2 = \sigma_y \otimes \sigma_y \otimes 1 \;, \quad
    \Gamma^3 = \sigma_z \otimes \sigma_y \otimes \sigma_y \;.
\end{eqnarray*}
The gauge field in this Majorana representation is $\mathcal{A
}_\mu = 1 \otimes 1 \otimes (A_\mu^{(-)} - A_\mu^{(+)} \sigma_y)$
where $A_\mu^{(\pm)} = \frac{1}{2}( A_\mu^{\vphantom{\rm T}} \pm
A_\mu^{\rm T})$ are the symmetric and skew parts of $A_\mu \in
\mathfrak{su}(N_c)$.

The operator $H = {\rm i}\Gamma^\mu (\partial_\mu - \mathcal{A}_
\mu)$ is imaginary skew, therefore ${\rm e}^{{\rm i}t H}$ is real
orthogonal.  This means that there exists a Nambu space
$\mathcal{V}$ with unitary structure $\langle \cdot , \cdot
\rangle$ and symmetric pairing $\{ \cdot , \cdot \}$, both of
which are preserved by the action of ${\rm e}^{{\rm i}t H}$.  No
change of physical meaning or interpretation is implied by the
identical rewriting from Dirac $D$ to Majorana $H$.  The fact that
Dirac fermions are not truly Majorana is encoded in a ${\rm U}
(1)$-symmetry $H {\rm e}^{{\rm i} \theta Q} = {\rm e}^{{\rm
i}\theta Q} H$ generated by $Q = 1 \otimes 1 \otimes \sigma_y$.

Now comes the essential point: since $H$ obeys $\bar H = - H$, the
chiral ``symmetry'' $H = - \Gamma_5 H \, \Gamma_5$ with $\Gamma_5
= 1 \otimes \sigma_x \otimes 1$ can be recast as a \emph{true}
symmetry:
\begin{displaymath}
    H= + \Gamma_5 \bar H\, \Gamma_5 = T H T^{-1} \;,
\end{displaymath}
with anti-linear $T : \Psi \mapsto \Gamma_5 \bar\Psi$.  Thus the
massless QCD Dirac operator is indeed associated with a symmetry
class in the present, post-Dyson sense: that's class $A$III,
realized by self-adjoint operators on Nambu space with Dirac ${\rm
U}(1)$-symmetry and an anti-unitary symmetry $T$.

\subsection{Classes $BD$I and $C$II}

Consider Hamiltonians $D$ still of the form (\ref{classAIII}) but
now with matrix entries taken from either the real numbers or the
real quaternions.  Their one-parameter groups ${\rm e}^{ {\rm
i}tD}$ belong to two further families of symmetric spaces:
\begin{displaymath}
\begin{array}{ll}
    \mbox{Class $BD$I}: &{\rm SO}(p+q) / {\rm SO}(p) \times
    {\rm SO}(q) \;, \\ \mbox{Class $C$II}: &{\rm USp}(2p+2q)
    / {\rm USp}(2p) \times {\rm USp}(2q) \;.
\end{array}
\end{displaymath}
These large families are known to be realized as symmetry classes
by the massless Dirac operator with gauge group ${\rm SU}(2)$ (for
$BD$I), or with fermions in the adjoint representation (for
$C$II). For the details we must refer to Verbaarschot's paper, as
there is no space left here.


\begin{thebibliography}{1}
%
\bibitem[1]{asz} Altland A, Simons BD, Zirnbauer MR (2002)
\emph{Theories of low-energy quasi-particle states in disordered
$d$-wave superconductors}; Phys.\ Rep.\ {\bf 359}, pp 283-354
%
\bibitem[2]{az} Altland A, Zirnbauer MR (1997) \emph{Non-standard
symmetry classes in mesoscopic normal-/superconducting hybrid
systems}; Phys.\ Rev.\ B {\bf 55}, pp 1142-1161
%
\bibitem[3]{dyson} Dyson FJ (1962) \emph{The threefold way: algebraic
structure of symmetry groups and ensembles in quantum mechanics};
J.\ Math.\ Phys.\ 3, pp 1199-1215
%
\bibitem[4]{helgason} Helgason S (1978) \emph{Differential
geometry, Lie groups and symmetric spaces}; Academic Press, New
York
%
\bibitem[5]{mehta} Mehta ML (1991) \emph{Random matrices}; Academic,
N.Y.
%
\bibitem[6]{weyl} Weyl H (1939) \emph{The classical groups: their
invariants and representations}; Princeton University Press
%
\bibitem[7]{verbaarschot} Verbaarschot JJM (1994) \emph{The spectrum
of the QCD Dirac operator and chiral random matrix theory: the
threefold way}; Phys.\ Rev.\ Lett.\ {\bf 72}, pp 2531-2533
%
\end{thebibliography}
\end{document}